\documentstyle[preprint,prl,aps]{revtex}
\begin{document}
\begin{titlepage}
\preprint{\vbox{\baselineskip 15pt{
\hbox{SISSA 197/94/EP}
\hbox{JINR -- E2 -- 94 -- 474}
\hbox{hep -- th/9412105}
\hbox{December  1994}
\hbox{}}}}
\title{ \large \bf
Different factorizations of the relativistic
finite -- difference Schroedinger equation and q-oscillators}
\author{R.M.Mir -- Kasimov\footnote{Permanent address:
Joint Institute for Nuclear Research,
Dubna, Moscow region,
141980, Russia \\
e -- mail: mirkr@thsun1.jinr.dubna.su  \\
Work supported in part by Grant INTAS
Ref. 9.3 -- 127
}}
\address{SISSA -- Scuola Internazionale Superiore di Studi Avanzati\\
Via Beirut 2-4, I-34013 Trieste, Italy}
\maketitle
\begin{abstract}
\begin{minipage}{5.4in}
\baselineskip 16pt
The concept of the one -- dimensional quantum mechanics in the
relativistic configurational space (RQM) is reviewed briefly. The
Relativistic Schroedinger equation
(RSE) arising here is the finite --
difference equation
with the step equal to the Compton wave
length of the particle.
The different generalizations of the Dirac -- Infeld-- Hall
factorizarion
method for this case are
constructed. This method enables us to find out all possible
finite-difference generalizations
of the most
important nonrelativistic integrable case -- the harmonic oscillator.
As it was shown in \cite{kmn},\cite{mir6} the case of RQM
the harmonic oscillator = $q$ -- oscillator.

It is also shown that the relativistic and nonrelativistic QM's
are different representations of the same theory. The
transformation connecting these two representations is found in explicit
form. It could be considered as the generalization of the
Kontorovich -- Lebedev transformation.
\end{minipage}
\end{abstract}
\end{titlepage}
\newpage

\leftline{\bf 1. Introduction. One -- dimensional
Relativistic Quantum Mechanics}

Referring to papers (\cite{sny} -- \cite{mir6}) where the concept of the
relativistic configurational space was introduced and developed,
we recall here only that
this concept is based on the simple fact that the mass shall of the
free
relativistic particle is the Lobachevsky space and the matrix
elements of
its group of motion could be considered as the
"relativistic plane waves".

In the one -- dimensional case the Lobachevsky space is a hyperbola:
\begin{equation}
\label{hyp2}p_0^2-p^2=1
\end{equation}
and the relativistic plane wave is the exponent
\begin{equation}
\label{pw2}
\begin{array}{c}
<x|p>=\left( p_0-p\right) ^{-ix}=e^{i\chi x} \\
-\infty <x<\infty
\end{array}
\end{equation}
where $x\ $is the relativistic coordinate, $\chi $ is rapidity:
\begin{equation}
\label{rap}
\begin{array}{c}
\begin{array}{c}
p_0=\cosh \chi ,\quad p=\sinh \chi \\
\chi =\ln \left( p_0+p\right) =rapidity,\qquad -\infty <\chi <\infty
\end{array}
\\
d\Omega _p=\frac{dp}{p_0}=d\chi
\end{array}
\end{equation}
We use the unit system, in which $ \hbar = c = m = 1 $.
The wave function $\psi \left( x\right) \ $ is connected
with its momentum
space counterpart by Fourier transformation
\begin{equation}
\label{fourier}\psi \left( x\right) =\frac 1{\sqrt{2\pi }}\int_{-\infty
}^\infty d\Omega _p<x|p>\psi \left( p\right)
\end{equation}
The plane waves obey the following orthogonality and
completeness conditions
$$
\begin{array}{c}
\frac 1{2\pi }\int_{-\infty }^\infty <x|p>d\Omega _p<p|x^{\prime }>=\delta
\left( x-x^{\prime }\right)
$$
\\
$$
\frac 1{2\pi }\int_{-\infty }^\infty <p|x>d\Omega _p<x|p^{\prime }>=\delta
\left( \chi -\chi ^{\prime }\right)
\end{array}
$$
The free Hamiltonian and the momentum are
a finite-difference operators:
$$
H_0=\cosh i\frac d{dx},\qquad p=-\sinh i\frac d{dx}
$$
The plane wave $<x|p>$ obeys the free  finite-difference RSE
equation:
\begin{equation}
\label{rsch3}\left( H_0-p_0\right) <x|p>=0
\end{equation}
We can write the Eq (\ref{rsch3}) in the ''nonrelativistic''
form, using the
simple relation of the hyperbolic trigonometry, the
"half--rapidity relation"
:
$$
\cosh \chi =1+2\sinh {}^2\frac \chi 2
$$
The operator of free relativistic kinetic energy takes a form
(we restore
for the time being the dimensional quantities):
\begin{equation}
\label{rham3}h_0=2mc^2\sinh {}^2\frac{i\hbar
}{2mc}\frac d{dx}=\frac{
\stackrel{\wedge }{k}^2}{2m}=H_0-mc^2
\end{equation}
where the corresponding momentum operator was introduced
\begin{equation}
\label{rmom}\stackrel{\wedge }{k}\ =-2mc\sinh \frac{i\hbar }{2mc}\frac
d{dx},\quad k=2mc\sinh {}\frac \chi 2
\end{equation}
The RSE takes the form:
\begin{equation}
\label{rsch4}
\left( h-e\right) \psi \left( x\right) =0,
\end{equation}
where
\begin{equation}
\label{rsch5}h=\frac{\stackrel{\wedge }{k}^2}{2m} +
V(x),\qquad e=\frac{k^2}2
\end{equation}

In this form RSE is indistinguishible with  nonrelativistic
Schroedinger equation, until the explicit form of the momentum operator
is fixed. Obviously in nonrelativistic limit we have:
\begin{equation}
\label{rsch6}
h \rightarrow -\frac{\hbar}{2m}\frac{d^{2}}{dx^{2}}
\end{equation}

One of the most important integrable cases of usual Schroedinger equation is
the linear oscillator problem. We shall construct the relativistic Hamiltonian
corresponding to  this important nonrelativistic problem
and see that there are several ways to do that. It is
impossible to introduce the notion of the elastic forces in the relativistic
configurational space. We shall proceed by analogy, trying to restore all
important features of the nonrelativistic quantum mechanical linear
oscillator properties:

\begin{enumerate}

\item
The integrability. In particular, this property of nonrelativistic
oscillator is reflected in factorizability of hamiltonian. This will be
exploited by us intensively.

\item
The correct nonrelativistic limit.

\item
The symmetry. We shall require the existence of both kinds of the
symmetry:the symmetry of the hamiltonian and the dynamical symmetry. Of
course it is obvious from the beginning that in the relativistic case these
symmetries must be in some sense generalized. We see that in our case the
''generalization'' means going over to the $q-$ deformed groups $SU_q\left(
2\right) $ and $SU_q\left( 1,1\right) $ with $q$ given by (\ref{q2}). By
other words the relativization = $q$-deformation.

\item
The existence of some generalization of the symmetry between the
coordinates and momenta, which in the nonrelativistic case follows from the
explicitly symmetric hamiltonian:
\end{enumerate}

\begin{equation}
\label{symh}H_{nonrelat}=\frac{P^2+Q^2}2
\end{equation}

\vskip 0.4cm
\leftline{\bf 2. The non -- relativistic factorization method.}

Let us consider the nonrelativistic one -- dimensional Hamiltonian
\begin{equation}
\label{nrsch1}
H=-\frac 12\frac{d^2}{dx^2}+V(x)
\end{equation}
with the positive -- definite ground state wave function
\begin{equation}
\label{nrsch2}
\psi _0\left( x\right) =e^{-\rho \left( x\right) }\geq 0
\end{equation}
and lowest energy level $E_0$
\begin{equation}
\label{nrsch3}
H\psi _0\left( x\right) =E_0\psi _0\left( x\right)
\end{equation}
We can express $V(x)$ in terms of $\rho \left( x\right) $ and $E_0:$
\begin{equation}
\label{nrsch4}
V(x)=\frac 12\left[ \left( \frac{d\rho \left( x\right) }{dx}\right)
 ^2-\frac{
d^2\rho \left( x\right) }{dx^2}\right] -E_0
\end{equation}
Now Hamiltonian can be written in the factorized form
\begin{equation}
\label{nrsch5}
H-E_0=a^{+}a^{-}
\end{equation}
where
\begin{equation}
\label{anr}
\begin{array}{c}
a^{\pm }=\frac 1{
\sqrt{2}}\left( \mp \frac d{dx}+\frac{d\rho \left( x\right) }
{dx}\right) =
\\ \mp \frac 1{\sqrt{2}}e^{\mp \rho \left( x\right) }\frac d{dx}
e^{\pm \rho
 \left( x\right) }
\end{array}
\end{equation}

The commutator of the the creation and annihlation operators
\begin{equation}
\label{comut}\left[ a^{+},a^{-}\right] =\frac{d^2\rho
\left( x\right) }{dx^2}
\end{equation}
is simply a function of $x$ and don't contain the
derivative operator as $
a^{+},a^{-}$ themselfs do. We stress that in general
case the commutator of
two first order differential operators is the
differential operator of the
second order. The absence of the derivatives in
the right hand side of (\ref
{comut}) is the consequence of the specific
construction (\ref{anr}) and of
course of the Leibnitz rule for differentiation operation:
\begin{equation}
\label{leib1}\frac d{dx}f_1\left( x\right)
\cdot f_2\left( x\right) =\left(
\frac d{dx}f_1\left( x\right) \right) \cdot
f_2\left( x\right) +f_1\left(
x\right) \cdot \left( \frac d{dx}f_2\left( x\right) \right)
\end{equation}

The harmonic oscillator case is singled out.
In this case the right hand
side of (\ref{comut}) is not an function of $x$, but a constant
$\omega $ ,
the frequency of oscillator:
\begin{equation}
\label{com1}\left[ a^{+},a^{-}\right] =const=\omega
\end{equation}
The $\rho \left( x\right) $ takes a form
\begin{equation}
\label{rho}\rho \left( x\right)
=\frac{\omega x^2}2=\frac{m\omega x^2}
{2\hbar }
\end{equation}
and
$a^{+},\,a^{-}$
are
\begin{equation}
\label{anr1}a^{\pm }=\frac 1{\sqrt{2}}\left( \mp
\frac dd+\omega x\right)
=\mp \frac i{\sqrt{2}}\left( \stackrel{\wedge
}{p}\pm i\omega x\right)
\end{equation}
The Hamiltonian $H$ could be written in different forms:
\begin{equation}
\label{oscham}
\begin{array}{c}
H=\frac 12\{a^{+},a^{-}\}=\alpha
a^{+}a^{-}+\left( 1-\alpha \right)
a^{-}a^{+}+\left( 2\alpha -1\right) E_0=a^{-}a^{+}-E_0 \\
E_0=\frac \omega 2,\qquad \alpha -arbitrary\ number
\end{array}
\end{equation}

The fact of integrability of the oscillator problem is reflected in
commutation relations of the creation and
annihlation operators with the
Hamiltonian:
\begin{equation}
\label{solv}\left[ H,a^{+}\right] =\omega
a^{+},\qquad \left[ H,a^{-}\right]
=-\omega a^{-}
\end{equation}
The relations (\ref{solv}) give us the simple
way for constructing the
spectrum and the eigenvectors of Hamiltonian $H.$
If $\Psi $ is an
eigenvector of $H$ $\left( H\Psi =E\Psi \right) $,
the functions $a^{+}\Psi $
and $a^{-}\Psi $ (provided that they are nonzero and
belong to $L_2\left(
R\right) $) and the new eigenvectors corresponding
to the eigenvalues $
E+\omega $ and $E-\omega $ respectively,
\begin{equation}
\label{solv1}
\begin{array}{c}
H\left( a^{+}\Psi \right) =a^{+}\left( H+\omega \right) \Psi =
\left(
E+\omega \right) a^{+}\Psi \\
H\left( a^{-}\Psi \right) =a^{-}\left( H-\omega \right) \Psi =
\left(
E-\omega \right) a^{-}\Psi
\end{array}
\end{equation}

In fact the integrability is not strictly
connected with commutators in
l.h. sides of (\ref{solv}). Suppose we have
instead of (\ref{solv}) some
other linear combination of $Ha^{+}$ and $a^{+}H$ of the form
\begin{equation}
\label{solv2}AHa^{+}-Ba^{+}H=\omega a^{+}
\end{equation}
Obviously this relation could be used for
the constructing the higher and
lover eigenvectors of the hamiltonian
in the same way as (\ref{solv}) was
used in (\ref{solv1}). Of course the
spectrum of the Hamiltonian will be
different. We shall have such an
situation when we consider the
factorization in the relativistic case.
The Leibnitz rule for the finite -
difference calculus is different from
(\ref{leib1}) and we shall see that
integrability of the relativistic
oscillator problem is provided by the
relation of the type (\ref{solv2}) rather than (\ref{solv1}).
Ultimately
this will lead us to the deformed symmetries for the relativistic
oscillator \cite{mir6}.

\vskip 0.4cm
\leftline{\bf 3. The relativistic factorization method.
The relativistic oscillator}

Let $\psi _0\left( x\right) $\ be\ the\
ground\ state\ wave\ function and $
e_0\ $ the lowest energy level in eq. (\ref{rsch4}). We
introduce the relativistic creation and annihilation
operators (\cite{mir6})
\begin{equation}
\label{A}A^{\pm }=\mp i\sqrt{2\ }\alpha \left( x\right) \
e^{\pm \rho \left(
x\right) }\
\sinh {}\frac i2\frac d{dx}\,\, e^{\mp \rho \left( x\right) }
\end{equation}
The difference between$\left( \ref{anr}\right)
$ and $\left( \ref{A}\right) $
is originated from the different Leibnitz rules
for the finite-difference
operators
as compared with the differential one
$\left( \ref{leib1}\right) $:
\begin{equation}
\label{leib2}
\begin{array}{c}
\sinh {}\frac i2\frac d{dx}\left( f_1\left( x\right)
\cdot f_2\left(
x\right) \right) =\left( \sinh {}\frac i2\frac
d{dx}f_1\left( x\right)
\right) \cdot \left( \cosh \frac i2\frac
d{dx}f_2\left( x\right) \right) +
\\
\left( \cosh \frac i2\frac d{dx}f_1\left(
x\right) \right) \cdot \left(
\sinh {}\frac i2\frac d{dx}f_2\left( x\right) \right) \\
\cosh \frac i2\frac d{dx}\left( f_1\left(
x\right) \cdot f_2\left( x\right)
\right) =\left( \cosh \frac i2\frac
d{dx}f_1\left( x\right) \right) \cdot
\left( \cosh \frac i2\frac d{dx}f_2\left( x\right)
\right) + \\
\left( \sinh {}\frac i2\frac d{dx}f_1\left( x\right)
\right) \cdot \left(
\sinh {}\frac i2\frac d{dx}f_2\left( x\right) \right)
\end{array}
\end{equation}
This new Leibnitz rules naturally suggest a new construction --
the
"$q\left( x\right)$ -mutator":
\begin{equation}
\label{q(x)-mu}
\begin{array}{c}
\left[ A^{-},A^{+}\right] _{q\left( x\right) } =
A^{-}\cdot e^{a\left(
x\right) }\cdot A^{+}-A^{+}\cdot e^{-a\left( x\right)
}\cdot A^{-} \\
q\left( x\right) =e^{a\left( x\right) }
\end{array}
\end{equation}
In the explicit form
\begin{equation}
\label{q(x)-mut}\left[ A^{-},A^{+}\right] _{q\left(
x\right) }=\frac{\alpha
\left( x\right) }2\left\{
\begin{array}{c}
e^{\frac i2\frac d{dx}}\cdot \alpha \left( x\right)
\cdot \sinh Z\left(
x\right) \cdot e^{\frac i2\frac d{dx}}+ \\
e^{-\frac i2\frac d{dx}}\cdot \alpha \left( x\right)
\cdot \sinh Z\left(
x\right) \cdot e^{-\frac i2\frac d{dx}}- \\
e^{\frac i2\frac d{dx}}\cdot \alpha \left( x\right)
\cdot \sinh \left(
Z\left( x\right) +2\rho _{\frac s2}\left( x\right)
\right) \cdot e^{-\frac
i2\frac d{dx}}- \\
e^{-\frac i2\frac d{dx}}\cdot \alpha \left( x\right)
\cdot \sinh \left(
Z\left( x\right) -2\rho _{\frac s2}\left( x\right)
\right) \cdot e^{\frac
i2\frac d{dx}}
\end{array}
\right\}
\end{equation}
where the following notations were introduced:
\begin{equation}
\label{not}
\begin{array}{c}
Z\left( x\right) = 2\rho \left( x\right) +a
\left( x\right) \ -2\rho _{\frac
c2}\left( x\right) \\
\rho _{\frac s2}\left( x\right) =
\sinh {}\frac i2\frac d{dx}\rho \left(
x\right) ,\quad \rho _{\frac c2}\left( x\right) =
\cosh \frac i2\frac
d{dx}\rho \left( x\right)
\end{array}
\end{equation}

The r.h.s. of $\left( \ref{q(x)-mut}\right) $ is
much more complicated than
the r.h.s. of the comutator of the nonrelativistic
creation and annihilation operators $a^{+}$ and
$a^{-}$ $\left( \ref{comut}
\right) $. We see that the condition that
there are no finite-difference
differentiations in the r.h.s. of
$\left( \ref{q(x)-mu}\right) $ is:
\begin{equation}
\label{not1}
Z\left( x\right) =0
\end{equation}
This gives us the relation connecting $q\left( x\right) $
 and $\rho \left(
x\right) $. For $ Z\left( x\right) =0 $  we have:
\begin{equation}
\label{q(x)-muta}\left[ A^{-},A^{+}\right] _{q
\left( x\right) }=-\alpha
\left( x\right) \cdot \sinh {}\frac i2\frac d{dx}\left[
\alpha \left(
x\right) \cdot \sinh \left( 2\rho _{\frac s2}\left( x\right)
\right) \right]
\end{equation}
Again, following the example of the nonrelativistic
oscillator we wonder
whether it is possible that the expression in the r.h.s. of
(\ref{q(x)-muta})
is equal to a constant as it was in the case of the
non-relativistic oscillator
(see $\left( \ref{com1}\right) $):
\begin{equation}
\label{q(x)-mutat}\left[ A^{-},A^{+}\right]
_{q\left( x\right)} = const
\end{equation}
This means that we shall consider
$\left( \ref{q(x)-mutat}\right) $ as an
equation for $\rho \left( x\right) $.
The remarkable fact is that this
problem is solvable, the solution coincides
with the nonrelativistic $\rho
\left( x\right) $ $\left( \ref{rho}\right) $
and obviously does not depend
on the velocity of light.

The factor
$\alpha \left( x\right) $
defined from
(\ref{q(x)-mutat})
is equal to
\begin{equation}
\label{alpha}\alpha \left( x\right) =
\frac 1{\cos \frac{\omega x}{2c}}
\end{equation}
The
$a\left( x\right) $
and
$q\left( x\right) $
(see \ref{q(x)-mu}) are now constants:
$$
\begin{array}{c}
a\left( x\right) =-\frac \omega 4 \\
q\left( x\right) =const=q=e^{-\frac \omega 4}
\end{array}
$$
or in dimensional units
\begin{equation}
\label{q2}q=e^{-\frac{\omega \hbar }{4mc^2}}
\end{equation}
and instead of the $q\left( x\right) -$mutator we have the
$q-$mutator
\begin{equation}
\label{q-mut1}\left[ A^{-},A^{+}\right]_q =
q\,A^{-}A^{+}-q^{-1}A^{\div }A^{-}
\end{equation}
and this $q-$mutator is equal to a constant
\begin{equation}
\label{q-mut2}\left[ A^{-},A^{+}\right] _q =
2\left( q^{-1}-q\right) = 4\sinh
\frac \omega 4
\end{equation}
The operators $A^{\pm }$ can be calculated now in explicit form
\begin{equation}
\label{A2}
\begin{array}{c}
A^{\pm }=\pm i
\sqrt{2}e^{\pm \frac \omega 8}\left( \sinh
{}\frac i2\frac d{dx}\mp i
\tan
\frac{\omega x}2
\cosh \frac i2\frac d{dx}\right) = \\ \mp \frac 1{\sqrt{2}%
}e^{\pm \frac{\omega x^2}2}\stackrel{\wedge }
{\cal D}e^{-\frac{\omega x^2}2}
\end{array}
\end{equation}
where
\begin{equation}
\label{D}\stackrel{\wedge }{\cal D}\ =
\ -\frac{2i}{\cos \frac{\omega x}2}
\cdot \sinh {}\frac i2\frac d{dx}
\end{equation}
The Hamiltonian of the relativistic oscillator
has the form \cite{mir6}:
\begin{equation}
\label{rhmi1}
\stackrel{\wedge }{h} = \frac{1}{2}\{ A^{-},\,A^{+}\}_{q} =
\frac{1}{2}\left( q\,A^{-}A^{+} + q^{-1}A^{+}A^{-} \right) =
2\left( \stackrel{\wedge }{T}^{2} - \cosh\frac{\omega}{4}\right)
\end{equation}
where
\begin{equation}
\label{T}
\stackrel{\wedge }{T} =
\frac{1}{\cos\frac{\omega x}{2}}\cosh\frac{i}{2}\frac{d}{dx} =
q^{-N-1/2}
\end{equation}
and operator $N$, the analog of the occupation number was introduced.
In \cite{mir6}
relations (\ref{T}) between Hamiltonian operator (\ref{rhmi1}),
finite-difference operator $\stackrel{\wedge}{T}$ and
the "occupation number operator"  $N$ were established.
It was also shown
(see also \cite{macf}) that
the eigenvalues of $N$ are equal to $ 0,1,2,... $.
The  $ q$-mutator of the
creation operator
$A^{+}$ with the Hamiltonian reflects the
integrability property of the
problem considered
\begin{equation}
\label{integrability}
\left[ A^{+},\stackrel{\wedge }{h} \right]_{q^{-1}} =
\left( q^{2} - q^{-2}\right)A^{+}
\end{equation}
The solution of the oscillator RSE
\begin{equation}
\label{rhmi2}
\stackrel{\wedge }{h}\,\Psi (x) = e\,\Psi\,(x)
\end{equation}
corresponding to the $n$-th energy level  $e = e_{n}$
has the form
\begin{equation}
\label{wfun1}
\Psi_{n}(x) = e^{-\frac{\omega x^{2}}{2}}\,h_{n}(x)
\end{equation}
where $h_{n}(x)$ are relativistic Hermite polynomials.
We refer the reader for the details
to \cite{mir6} and write down here
only the
necessary relations for the $h_{n}$.
\begin{itemize}
\item
The Rodriduez formula:
\begin{equation}
\label{Rodrig}
h_{n+1}(x) = \stackrel{\wedge }{R}\,h_{n}(x) =
\left( \frac{1}{\sqrt{\omega}} \right)
\,e^{\omega x^{2}}
\,\stackrel {\wedge}{\cal D}\,e^{-\omega x^{2}}\,h_{n}(x)
\end{equation}
It could be easily seen that operator
$\stackrel{\wedge}{R}$ acting on {\em arbitrary}
polynomial of $n-th$ degree gives an polynomial
of $n+1 \, - th$ degree  \\
\item
The finite -- difference equation:
\begin{equation}
\label{T1}
 \stackrel{\wedge }{K}\, h_{n}(x) =
e^{\frac{(n+1/2)\omega}{4}}\,h_{n}(x)
\end{equation}
where
\begin{equation}
\label{K}
 \stackrel{\wedge }{K} =
e^{\frac{\omega x^{2}}{2}}\,\hat{T}\,
e^{-\frac{\omega x^{2}}{2}}
\end{equation}
\end{itemize}

\skip 0.4cm
\leftline{\bf 4. The inhomogeneous lattice, double periodicity}
{\bf of relativistic Hamiltonian and orthogonality conditions}

We defined the factor $\alpha \left( x\right) $
as the solution of the equation
$\left( \ref{q(x)-mutat}\right) $,
but it is important to stress that it coincides
with the characteristic lattice factor coming from the general
theory of polynomials of the discrete variables (see \cite{sus1}).
It was shown in the theory of relativistic oscillator \cite{mir6}
that the wave
functions,
as the solutions of the finite-difference RSE,
are expressed in terms of the relativistic Hermite polynomials
with characteristic dependence on the
lattice argument  $\sin\frac{\omega x}{2} $.
{}From the other side  it was stressed in (\cite{mir6}) that
$h_{n}$-s are $q$-generalizations of the usual Hermite
polynomials
(see \cite{ask},\cite{rahm},\cite{at} and references therein).
The explicit form of the natural argument of the relativistic
Hermite polynomials is defined by the lattice. By other words,
it is the lattice, which indicates the specific change of
variables
in finite-difference RSE for generalized Hermite polynomials.

In usual differential calculus, changing the argument of the function
\begin{equation}
\label{change1}
f(x) = f\left( s(x) \right)
\end{equation}
we have the following formula for the derivative in new argument:
\begin{equation}
\label{change2}
\frac{df}{ds} = \frac{\frac{df}{dx}}{\frac{ds}{dx}}
\end{equation}
Analogous formula for finite-difference differentiation
has the form:
\begin{equation}
\label{change3}
\sinh \frac{i}{2}\frac{d}{dx} f(x) =
\frac{\sinh  \frac{i}{2}\frac{d}{dx} f\left(
s(x)\right)}{\sinh\frac{i}{2}\frac{d}{dx}s(x)}
\end{equation}
Taking into account that
\begin{equation}
\label{change4}
\sinh \frac{i}{2}\frac{d}{dx} \sin\frac{\omega x}{2}
= i \sinh\frac{\omega}{4}\cos\frac{\omega x}{2}
\end{equation}
we convince that (\ref{D}) is indeed the finite-difference
differentiation operator in terms of lattice variable
$\sin\frac{\omega x}{2}$.

Another important feature of RSE is that in the theory
of relativistic
oscillator the lattice factor
$\alpha \left(x\right) $
is also connected with the cutting factor of
the oscillatory wave
function or the ground state wave function
$e^{-\frac{\omega x^2}2}$. We
see
\begin{equation}
\label{change5}
\cosh \frac {i}{ 2}\frac d{dx}e^{-\frac{\omega x^2}2} =
e^{\frac \omega 8}
\cos \frac{\omega x}2
e^{-\frac{\omega x^2}2}=\frac 1{\alpha \left( x\right) }e^{-
\frac{\omega x^2}2}
\end{equation}
We see now clearly that lattice plays nontrivial role in
the finite-difference relativistic quantum mechanics.
But this is only the beginning of the story.

Maybe the most  difficult problem connected with
this change of variables
in finite-difference RSE is to understand
the role which plays the periodicity of the argument of relativistic
Hermite polynomials
$\sin\frac{\omega x}{2}$.
In particular, this problem is reflected in
the orthogonality condition for the wave functions
of the relativistic
oscillator (\ref{wfun1}).

This orthogonality condition is written as
\begin{equation}
\label{ort1}
\int_{-\infty}^{\infty}
h_{n}(x)\, h_{m}(x)\,
\mu_{1}(x)
dx  = \delta_{nm} J_{n}
\end{equation}
where
\begin{equation}
\label{meas1}
\mu_{1}(x) = e^{-\omega x^{2}}\,
\cos \frac{\omega x}{2}
\end{equation}
(see the expression for $J_{n}$ and other details in \cite{mir6} ).
In fact the equation (\ref{ort1}) is not the orthogonality
relation, because the the quantity
$  \cos \frac{\omega x}{2} $ is not positively definite and
could not be considered as a measure.
At the same time the consideration of the infinite interval is
necessary from the physical point of wiev, because this interval
is one -- dimensional physical space in which the Schroedinger
equation is written.
There is obvious contradiction between
the probabilistic interpretation
of the wave function in full physical space and the explicit
form of the exact solutions of the Schroedinger equation.
{}From the other side in the theory of the q -- polynomials
( see \cite{ask},
\cite{rahm}, \cite{at} and references therein) the
orthogonality condition is written as integral over the finite
interval $ (- \frac{\pi}{\omega}, \frac{\pi}{\omega}) $, (where
$  \cos \frac{\omega x}{2} $  is positive definite)
\begin{equation}
\label{ort2}
\int_{-\frac{\pi}{\omega}}^{\frac{\pi}{\omega}}
h_{n}(x)h_{m}(x)\,
\mu_{2}(x)\,dx  = \delta_{nm}\, J_{n}
\end{equation}
with different
well defined measure:
\begin{equation}
\label{meas2}
\mu _{2}(x) =
\, \Theta_{4} \left( 2\pi i x, e^{-\frac{4\pi^{2}}{\omega}} \right)\,
e^{-\omega x^{2}}\,
\cos \frac{\omega x}{2} =
 \, \Theta_{4} \left( 2\pi i x, e^{-\frac{4\pi^{2}}{\omega}} \right)\,
\mu _{1}(x)
\end{equation}
where $ \Theta_{4} \left( 2\pi i x,
e^{-\frac{4\pi^{2}}{\omega}} \right) $
is the
{\em Theta -- function} \cite{whit}:

Trying to understand the role of the inhomogeneous lattice
$s(x) = \sin\frac{\omega x}{2} $, the nature of the  change
of variables (\ref{change1}),
and the related problem of the ortoghonality condition for the
relativistic Hermite polynomial, we consider the periodicity
properties of the Hamiltonian  (\ref{rhmi1}) for the
relativistic oscillator and the corresponding
properties of the oscillator wave functions.
The Hamiltonian (\ref{rhmi1}) has two different periodicity
properties:

\begin{enumerate}
\item
{\em The first periodicity property.} \\
The step of finite -- difference RSE (\ref{rsch4}), (as well as
the step in the free equation (\ref{rhmi2})) is
equal to $\frac{i}{2}$.
This means that any function $\alpha (x)$, periodical with
the period $\frac{i}{2}$ i.e.
\begin{equation}
\label{period1}
\alpha \left( x+\frac{i}{2}\right) = \alpha \left( x \right)
\end{equation}
 "is a constant in respect to RSE". By other words, if $\psi (x)$
is the solution of RSE, then
$\alpha(x)\,\psi (x)$ is also the solution
of the same equation.
 The step of difference operators
$\stackrel {\wedge}{\cal D}$ (\ref{D}),
$\stackrel{\wedge }{T}$
(\ref{T})  and
$\stackrel{\wedge }{K}$ (\ref{K})
also equals $ \frac{i}{2} $\\

\item
{\em The second periodicity property.} \\
The Hamiltonian (\ref{rhmi1}) is periodic operator, with
the period $\frac{2\pi}{\omega}$:
\begin{equation}
\label{period2}
 \stackrel{\wedge }{h} \left( x + \frac{2\pi}{\omega}\, ,
\cosh\frac{i}{2}\frac{d}{dx}  \right) =
 \stackrel{\wedge }{h} \left( x \, ,
\cosh\frac{i}{2}\frac{d}{dx}  \right)
\end{equation}

The operators
$\stackrel {\wedge}{\cal D}$ (\ref{D}),
 $\stackrel{\wedge}{T}$
(\ref{T})
and $\stackrel{\wedge}{K}$ (\ref{K})  are quasi -- periodic
in respect to shifts
by $ \,\frac{2 \pi}{\omega} $:
\begin{equation}
\label{period3}
 \stackrel {\wedge}{\cal D} \left( x+ \frac{2 \pi}{\omega},
\sinh\frac{i}{2}\frac{d}{dx} \right)
  =
- \stackrel{\wedge}{\cal D} \left( x,
\sinh\frac{i}{2}\frac{d}{dx} \right)
\end{equation}
\begin{equation}
\label{period4}
  \stackrel {\wedge}{T}\left(x+ \frac{2 \pi}{\omega},
\cosh\frac{i}{2}\frac{d}{dx} \right)  =
-  \stackrel {\wedge}{T}\left(x,
\cosh\frac{i}{2}\frac{d}{dx} \right)
\end{equation}
\begin{equation}
\label{period5}
 \stackrel{\wedge }{K}\left(x+ \frac{2 \pi}{\omega},
\cosh\frac{i}{2}\frac{d}{dx} \right)  =
 \stackrel{\wedge }{K}\left(x,
\cosh\frac{i}{2}\frac{d}{dx} \right)
\end{equation}
These periodicity properties of
the oscillator RSE  are very important.
They actually define the characteristic inhomogeneous lattice
$s(x) = \sin\frac{\omega x}{2}$, the specific dependence
 on this variable of the relativistic Hermite polynomials, the
form of orthogonality conditions described earlier and the
connection of the theory considered with $\Theta$ --
functions. \\

\end{enumerate}

The periodicity properties of the Hamiltonian are reflected in
the measures $\mu_{1}(x)$ and $\mu_{2}(x)$ and relativistic
Hermite polynomials $h_{n}(x)$:

\begin{enumerate}
\item